\documentclass[%
 aip,
 amsmath,amssymb,
preprint,%
]{revtex4-1}

\usepackage{graphicx}
\usepackage{dcolumn}
\usepackage{bm}

\usepackage[utf8]{inputenc}
\usepackage[T1]{fontenc}
\usepackage{mathptmx}
\usepackage{etoolbox}

\usepackage{subfigure}
\usepackage{xparse}

\newcounter{subfigs}[figure]

\makeatletter
\NewDocumentCommand{\subfigimg}{s O{} m D<>{10pt} O{2\baselineskip} m}{%
    \IfBooleanTF{#1}%
        {\@subfigimg{#2}{#3}{#6}{0}{#4}{#5}}%
        {\@subfigimg{#2}{#3}{#6}{1}{#4}{#5}}%
}
\newcommand*{\@subfigimg}[6]{%
  \bgroup%
  \advance\c@figure by #4%
  \refstepcounter{subfigs}%
  \ifx\relax#3\relax\else%
  \label{#3}%
  \fi%
  \setbox1=\hbox{\includegraphics[#1]{#2}}
  \leavevmode\rlap{\usebox1}
  \rlap{\hspace*{-7.5pt}\raisebox{\dimexpr\ht1-#6\relax}{(\alph{subfigs})}}
  \phantom{\usebox1}
  \egroup%
}

\usepackage{xr}
\usepackage{amsmath}
\usepackage{xcolor}
\usepackage{epstopdf}
\usepackage{color}
\usepackage{multirow}

\makeatletter
\def\@email#1#2{%
 \endgroup
 \patchcmd{\titleblock@produce}
  {\frontmatter@RRAPformat}
  {\frontmatter@RRAPformat{\produce@RRAP{*#1\href{mailto:#2}{#2}}}\frontmatter@RRAPformat}
  {}{}
}%
\makeatother
\begin{document}

\preprint{}

\title[]{Time Crystals transforming Frequency Combs in Tunable Photonic Oscillators}
%
\author{Georgia Himona}
 \affiliation{School of Applied Mathematical and Physical Sciences, National Technical University of Athens, Athens 15780, Greece}%

\author{Vassilios Kovanis}%
 \affiliation{Bradley Department of Electrical and Computer Engineering, Virginia Tech Research Center, Arlington, Virginia 22203, USA
}%

\author{Yannis Kominis}
\affiliation{School of Applied Mathematical and Physical Sciences, National Technical University of Athens, Athens 15780, Greece}%
\email{gkomin@central.ntua.gr}
\date{\today}

\begin{abstract}
The response of a tunable photonic oscillator, consisting of an Optically Injected Semiconductor Laser, under an injected Frequency Comb is considered with the utilization of the concept of the Time Crystal that has been widely used for the study of driven nonlinear oscillators in the context of mathematical biology. The dynamics of the original system reduce to a radically simple one-dimensional circle map with properties and bifurcations determined by the specific features of the Time Crystal fully describing the phase response of the limit cycle oscillation. The circle map is shown to accurately model the dynamics of the original nonlinear system of ordinary differential equations and capable for providing conditions for resonant synchronization resulting to output frequency combs with tunable shape characteristics. Such theoretical developments can have potential for significant photonic signal processing applications.
\end{abstract}

\maketitle


\section{\label{sec:introduction} Introduction}
Periodic forcing of nonlinear oscillators has been a topic of broad interest to scientists, engineers and mathematicians, as it leads to many different types of dynamics as well as interesting technological applications. Typical dynamical behaviors include the entrainment of the forced oscillator to the periodic input, complicated patterns of synchronization between the periodic input and the forced oscillator, quasi-periodic dynamics, chaos, and multistability, occurring in a vast variety of physical and man-made systems ranging from  cardiac and circadian rhythms of humans and animals to mechanical, electrical and optical systems \cite{Pikovsky}. Relevant technological applications range from electronic phase-locked loops \cite{Adler} to microwave photonics applications for broadband telecommunications, arbitrary waveform generation, sensor networks and instrumentation \cite{Yao2009, Qi2011},  based on photonic integrated circuits consisting of coupled semiconductor lasers \cite{Erneux_book, Ohtsubo_13}.

The nonlinear oscillators support self-sustained periodic dynamics corresponding to stable limit cycles of the underlying model. The limit cycles are characterized by their frequencies and oscillation amplitudes in all the components of the system as well as by their spectral content, since they are in general non-harmonic. Moreover, a stable limit cycle is characterized by the structure of its \emph{Isochrons} that partition its basin of attraction with respect to the \emph{asymptotic phase} of each initial condition \cite{Winfree1974, Guckenheimer, Winfree}. The latter uniquely determines the phase response of the limit cycle under perturbations due to either external driving or coupling to other oscillators \cite{Pikovsky}. The phase response is described by the \emph{Phase Transition Curves} (PTCs) or the \emph{Phase Response Curves} (PRCs) of a stable limit cycle, originally defined in the context of nonlinear dynamics in mathematical biology \cite{Winfree, Izhikevich}. The concept of the \emph{Time Crystal} has been originally introduced by A. Winfree \cite{Winfree} in the 1970's in order to provide a compact form containing all the phase response information of a stable limit cycle. The time crystal describes the new asymptotic phase ($\phi$) of the system as a function of the phase ($\theta$) and the amplitude ($A$) of a perturbation, and its name reflects its time-periodic character with respect to the original and the new phase of the system. It is worth mentioning that the concept of the time crystal has been re-introduced in the physics literature in 2010's by F. Wilczek for quantum \cite{Wilczek2012} and classical \cite{Shapere2012} systems. 

\begin{figure*}
\begin{center}
    \includegraphics[width=1\columnwidth]{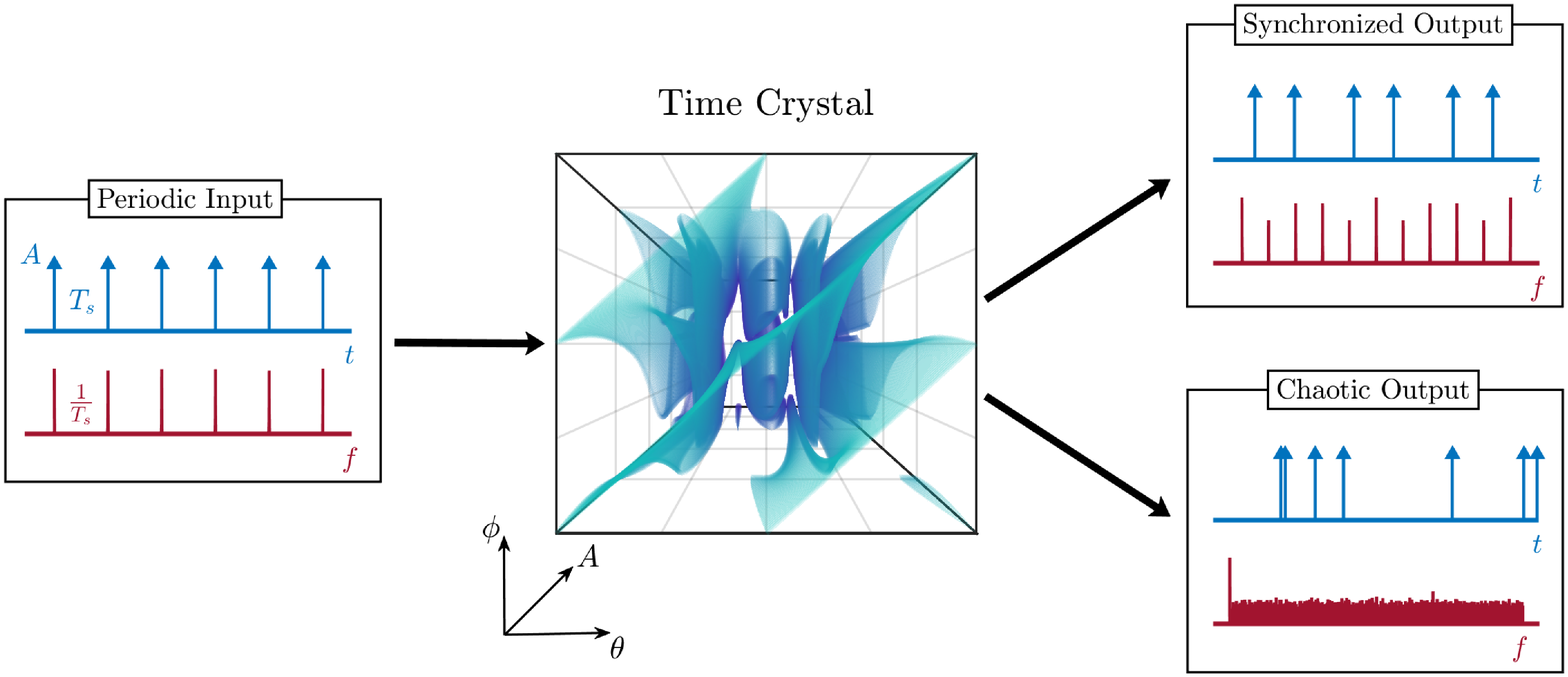}
    \caption{A time crystal fully characterizes the phase response of an oscillator to periodic input in the form of the frequency comb. Depending on the features of the time crystal as well as the amplitude $(A)$, frequency $(T_s)$ and input phase $(\theta)$ of the periodic forcing, the output phase $(\phi)$ series may correspond either to a resonantly synchronized or to a chaotic output. The time crystal essentially transforms an input frequency comb either to another comb, with controllable line spacing, or to a continuous spectrum.} 
    \label{fig_concept}
\end{center}
\end{figure*}

In a tunable oscillator, the limit cycle characteristics, namely its amplitude, frequency content and phase response, as described by the associated time crystal, can be parametrically modified. The latter also suggest a parametric modification of the qualitative features of the output of the system under a periodic pulsatile stimulation in the form of a time (and frequency) comb. In that sense, the time crystal, as the phase response function of the system, transforms a frequency comb input of the oscillator to provide its output, serving as a transfer function, as shown in Fig. \ref{fig_concept}. Due to the nonlinear character of the oscillation and its corresponding complex isochron structure (described by the time crystal), the output of the system can have radically different qualitative features, crucially depending on the features of the time crystal. The output of the system can range from bounded chaotic oscillations, having a continuous frequency spectrum, to periodic oscillations, having a discrete frequency spectrum consisting of equidistant spectral lines. The latter corresponds to various resonances in the synchronization between the external driving frequency comb and the limit cycle and, in that case, the input frequency comb is shaped to a different output frequency comb. The unique usefulness of the information contained in the time crystal of the system is related to the fact that it can be used to accurately model the complex synchronization dynamics of the system under a periodic pulsatile stimulation input by the iteration of a one-dimensional circle map, resulting in a significant dimensional reduction that enables the systematic investigation of the synchronization properties and the parametric control of the system's output. 

In this paper, we consider a fundamental tunable photonic oscillator consisting of an \emph{Optically Injected Semiconductor Laser} (OISL) where two semiconductor lasers are coupled in a master-slave configuration. The system is well-known for supporting the existence of robust stable limit cycles with multi-harmonic spectral content \cite{Kovanis2019} as well as a rich set of complex dynamical features \cite{Phys_Rep}, and it is intensively studied both theoretically and experimentally \cite{Kovanis_95a, Kovanis_95b} due to its technological significance \cite{Qi2011, Valagiannopoulos2021}. Recently, an OISL with a modulated optically injected signal, in the form of a frequency comb, has been considered as a tunable signal-processing unit \cite{Desmet2020, Schleimer2009, Shields2022} capable of providing a large variety of outputs with desired spectral features. The periodic modulation of the injected signal has been shown to drive self-sustained periodic oscillations excited from oscillation relaxation oscillations to a phase-locked state, corresponding to a fixed point of the underlying model, under appropriate conditions for frequency detuning with respect to the relaxation oscillation frequency of the modulated injected signal \cite{Lin2009, Gavrielides2014, Shortiss2019, Lingnau2020, Doumbia2020a, Doumbia2020b}. It is worth emphasizing that the complex dynamics of a system under periodic pulse modulation is essentially different under equilibrium (fixed point) and oscillatory (limit cycle) operation. In this work, we investigate the case where the periodically modulated injected signal is applied to an OISL in the parameter range where a self-sustained oscillation, corresponding to a stable limit cycle, already exists \cite{AlMulla2018, AlMulla2020, Iso22} even without injection modulation, and we study the role of its phase response, as described by its time crystal, in shaping its output.

\section{The Time Crystals of an Optically Injected Semiconductor Laser}
\subsection{Model}
The fundamental model describing the dynamics of the normalized complex electric field $Y=x+iy$ and the normalized excess carrier density $Z$ in an Optically Injected Semiconductor Laser is 
\begin{equation}
    \renewcommand{\arraystretch}{1.5}
    \begin{array}{l}\dfrac{dx}{dt}=(x-\alpha y)Z+\Omega y+\eta\\
    \dfrac{dy}{dt}=(y+\alpha x)Z-\Omega x \label{oisl}\\
    T\dfrac{dZ}{dt}=P-Z-(1+2Z)(x^2+y^2)
    \end{array} 
\end{equation}
where time has been rescaled to the photon lifetime $\tau_p$ -typically measured in ps-, $\alpha$ is the linewidth enhancement factor, $T$ is the ratio of carrier to photon lifetimes, and $P$ is the normalized excess electrical pumping rate of the slave laser.  $\eta$ and $\Omega$ correspond to the normalized injection rate and the normalized detuning between the frequency of the master laser and the frequency of the free-running slave laser, respectively. Characteristic time scales of the system are related to the damping rate $\gamma_R=(1+2P)/2T$ and the frequency $\omega_R=\sqrt{2P/T}$ of the free running laser relaxation oscillations \cite{Erneux_book, Phys_Rep, Valagiannopoulos2021, Herrera2022}.

\subsection{Time Crystals}
\begin{figure}
\begin{center}
    \subfigimg[width=0.45\columnwidth]{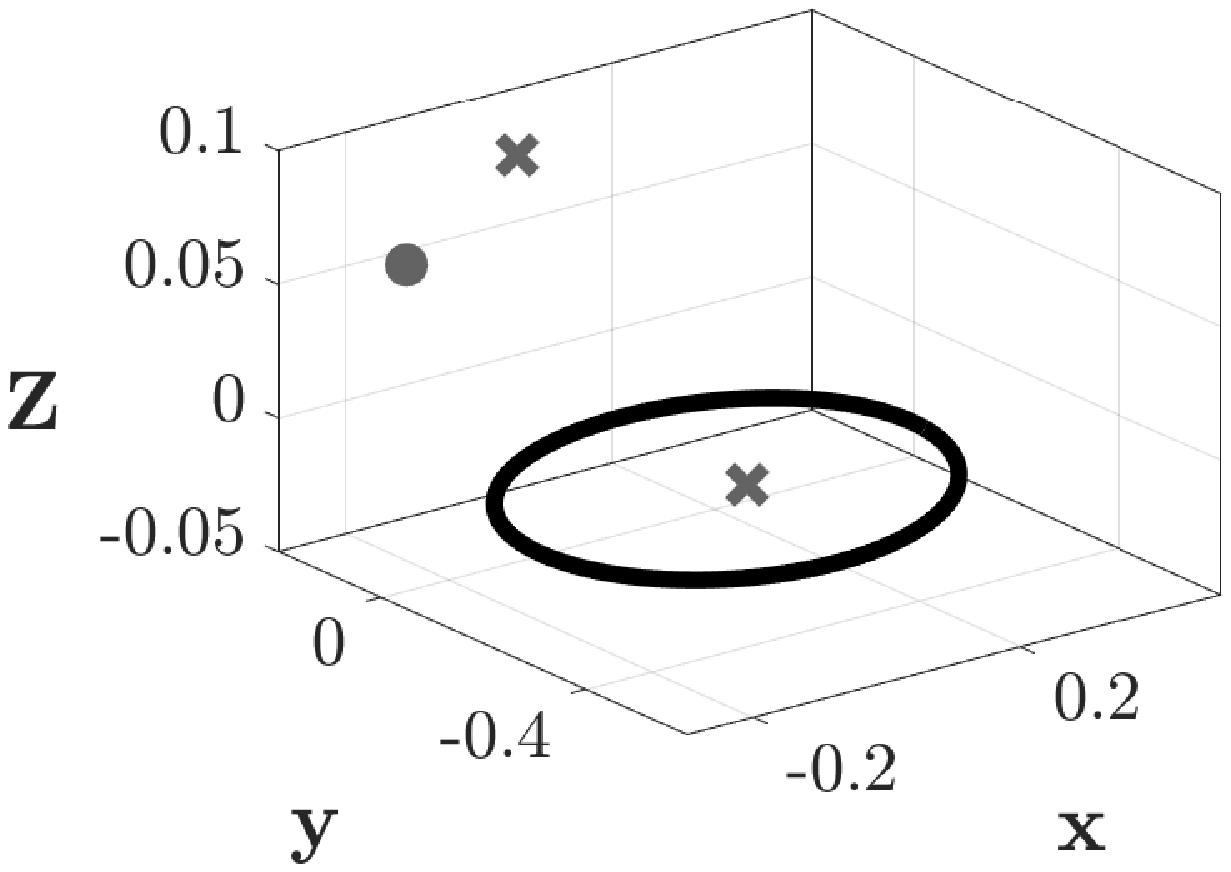}[1\baselineskip]{ps2}
    \subfigimg[width=0.45\columnwidth]{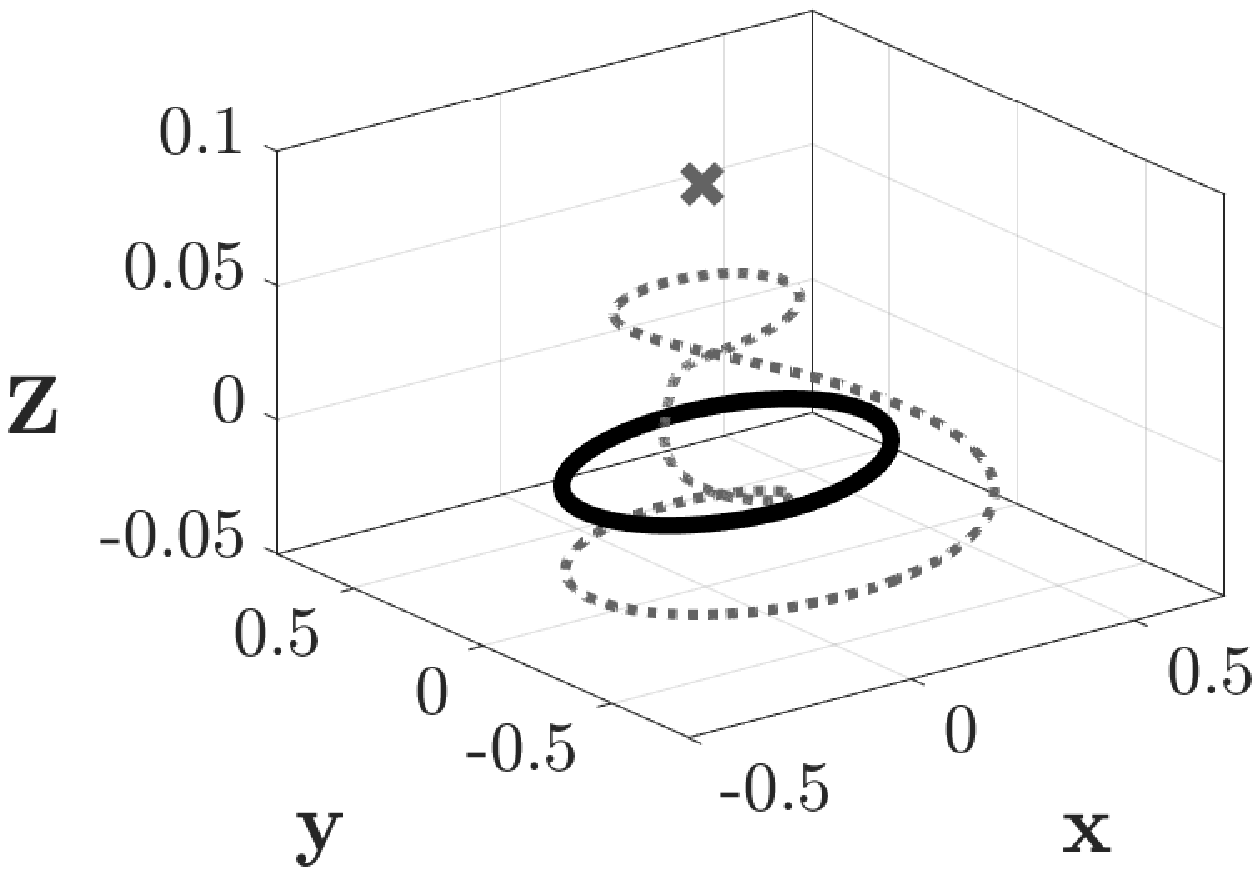}[1\baselineskip]{ps1}
    \caption{Phase space of the system (\ref{oisl}) for (a) $(\Omega, \eta)= (0.06,0.02)$ and (b) $(\Omega, \eta)= (0.06,0.0025)$.  x-points denote saddle-foci, the $\bullet$ point denotes an unstable focus-node, solid and dotted curves denote stable and unstable limit cycles, respectively. Period of limit cycle in (a) $T_0=63.5$, and in (b) $T_0=99.5865$. The unstable limit cycle along with the stable manifolds of the saddles determine the phaseless sets of the stable limit cycles.} 
    \label{phase space}
\end{center}
\end{figure}

For each set of parameters values in Fig. \ref{phase space}, system (\ref{oisl}) exhibits an attracting limit cycle of period $T_0$, or by renormalization, of period $1$; thus its dynamics is dissipative and can be described efficiently by the introduction of an \textit{asymptotic phase} \cite{Guckenheimer, Winfree, Pikovsky, Izhikevich}. Any initial condition within the basin of attraction of the limit cycle can be labeled with an asymptotic phase that is the relative phase translation with which the system returns to the limit cycle after being perturbed to the specific initial condition due to an external stimulation. The \textit{phaseless set} of the limit cycle corresponds to the complement of its basin of attraction with respect to the total phase space of the system and consists of other possibly coexisting stable fixed points and limit cycles with their basins of attraction, or stable manifolds of saddles. Through the application of an identical stimulation to every point on the stable limit cycle of the system an \textit{initial condition curve} \cite{Winfree, Campbell} is obtained. The measurement of the induced phase difference between the limit cycle and the initial condition curve leads to the definition of the \textit{Phase Response Curve}, 
\begin{equation*}
\text{PRC}(\theta)=\theta_{new}-\theta
\end{equation*}
or equivalently, the \textit{Phase Transition Curve}, 
\begin{equation*}
\text{PTC}(\theta)=[\theta+\text{PRC}(\theta)]\mod1.
\end{equation*}

\begin{figure}
\begin{center}
    \subfigimg[width=0.9\columnwidth]{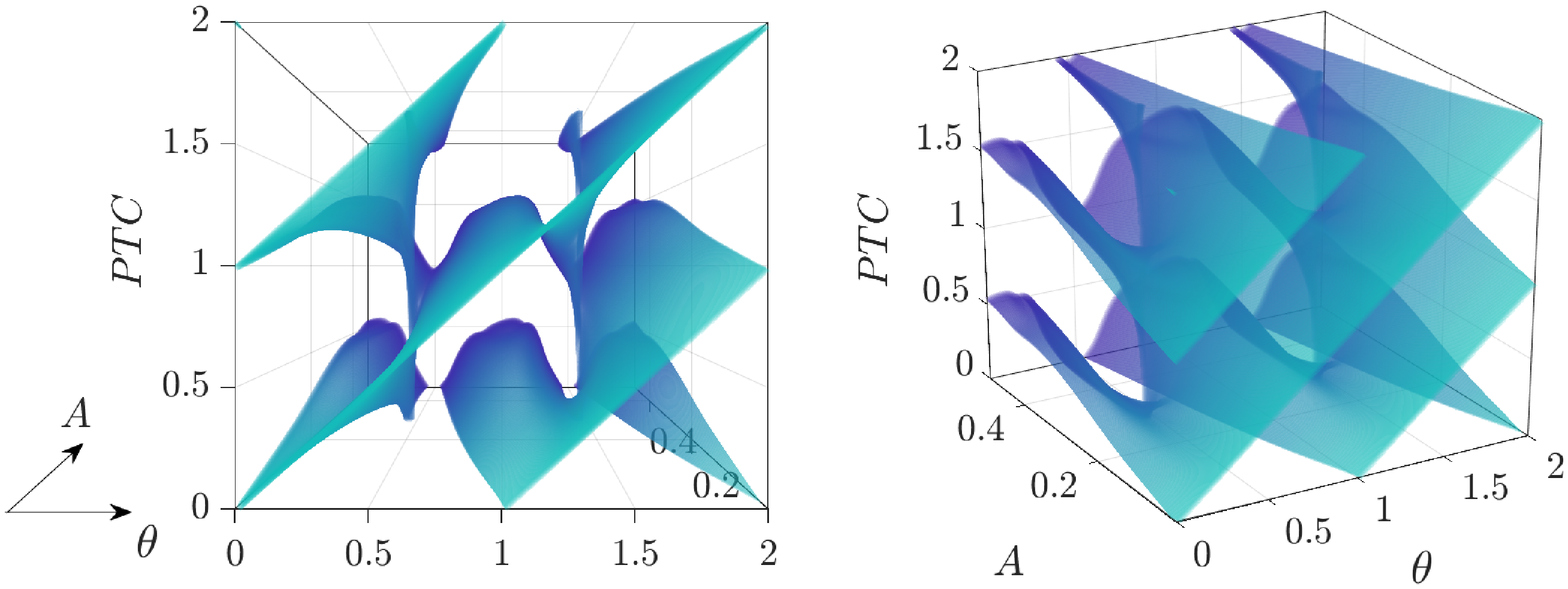}[1\baselineskip]{tc2} 
    \subfigimg[width=0.9\columnwidth]{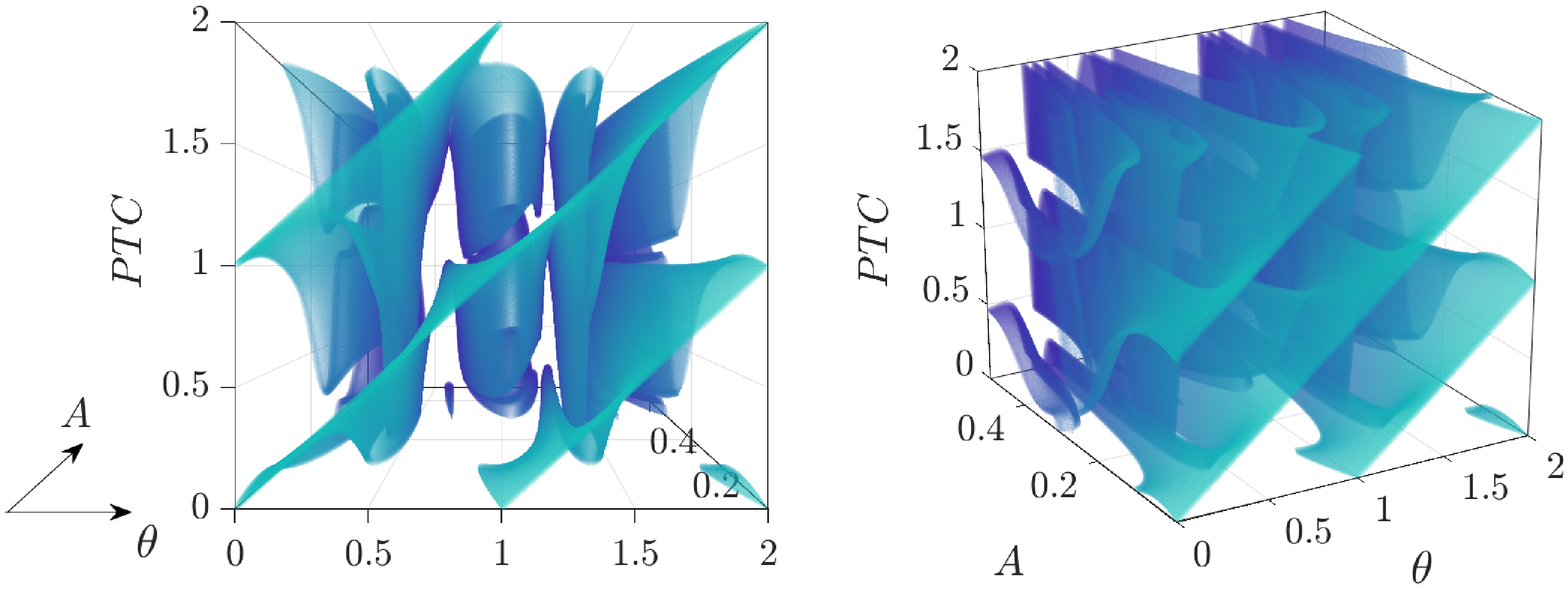}[1\baselineskip]{tc1}
    \caption{Time crystals corresponding to the limit cycles of system (\ref{oisl}) for $(\Omega, \eta)= (0.06,0.02)$ (a) and $(\Omega, \eta)= (0.06,0.0025)$ (b). $\theta\in[0,2)$ and $A\in[0,0.5]$.} 
    \label{time crystal}
\end{center}
\end{figure}

\begin{figure}
\begin{center}
    \subfigimg[width=0.495\columnwidth]{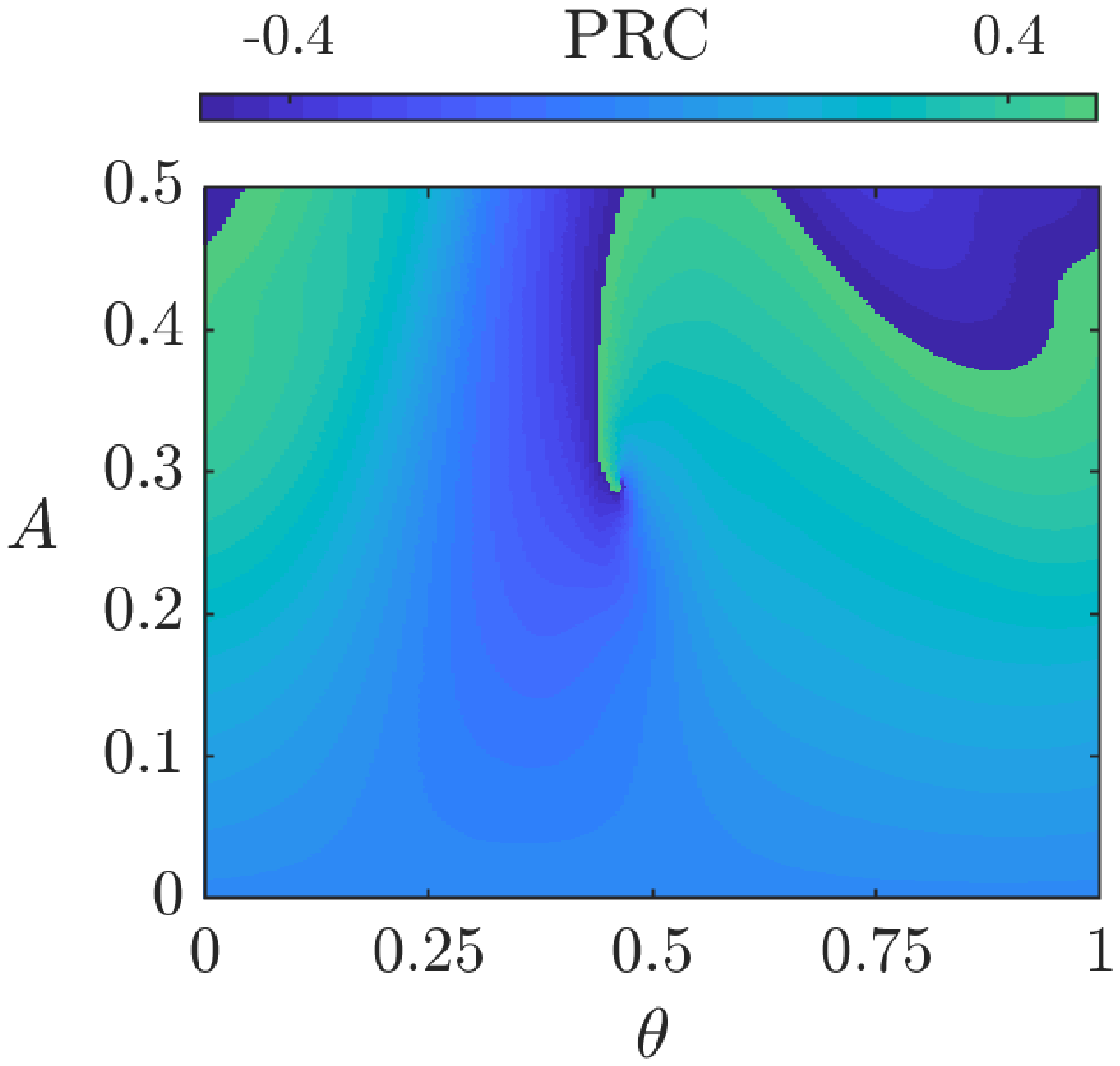}[1\baselineskip]{gprc2} 
    \subfigimg[width=0.495\columnwidth]{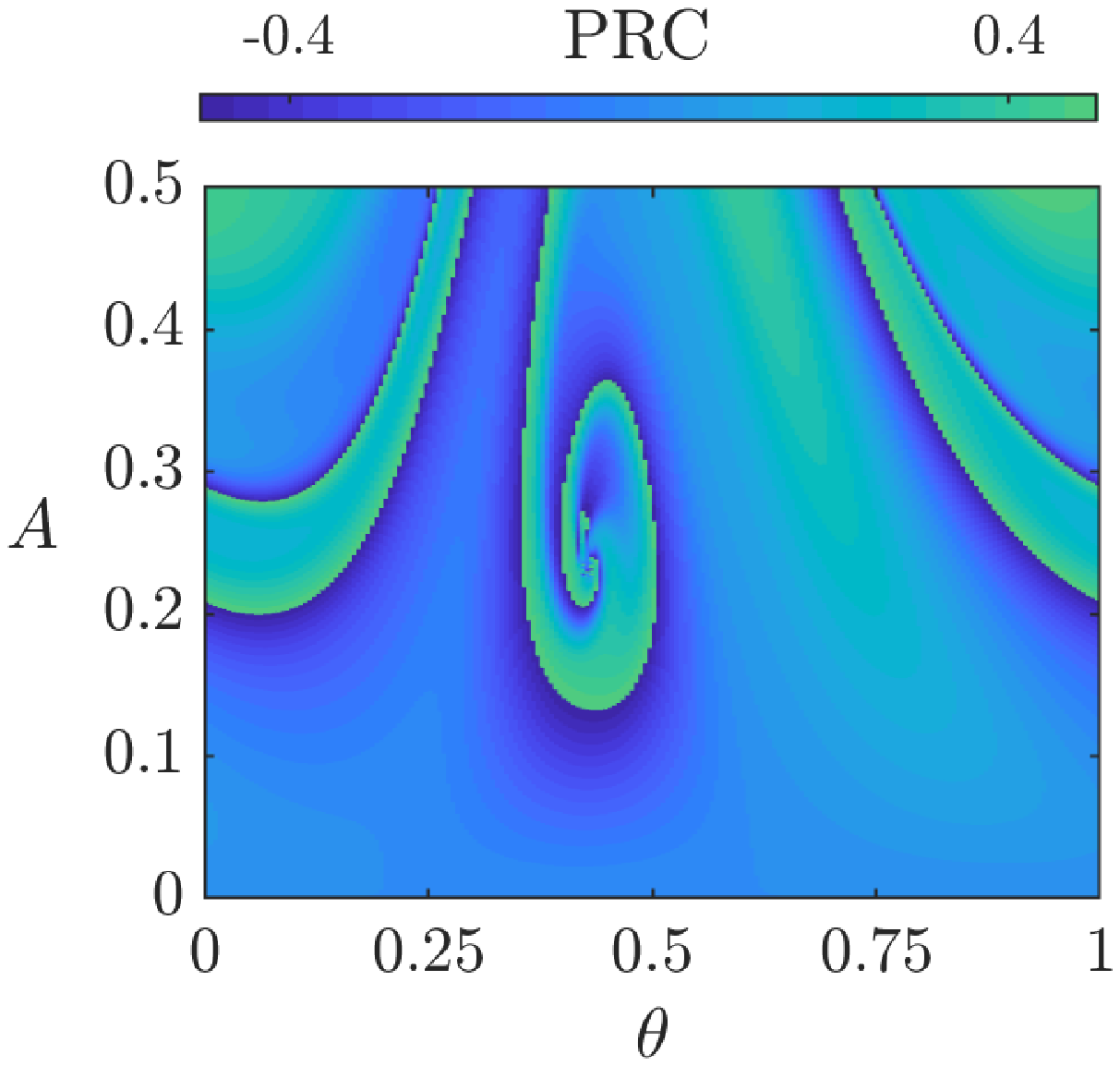}[1\baselineskip]{gprc1}
    \caption{Generalized PRCs corresponding to the limit cycles of system (\ref{oisl}) for $(\Omega, \eta)= (0.06,0.02)$ (a) and $(\Omega, \eta)= (0.06,0.0025)$ (b). $\theta\in[0,1)$ and $A\in[0,0.5]$.} 
    \label{GPRC}
\end{center}
\end{figure}

Considering the Phase Transition Curve also as a function of the amplitude $A$ of the applied stimulus, a $2-$dimensional surface can be constructed, as shown in Fig. \ref{time crystal}, where the $(A, \theta)$ constitute the stimulus plane and PTC is the phase transition representation. This surface is periodic along $\theta$ and $\text{PTC}$ coordinates, both representing time in units of the period $T_0$, resembling a $2-$dimensional crystal lattice, consequently defined as a \textit{Time Crystal} by Winfree \cite{Winfree}. The discontinuities of the time crystal imply change of the resetting type; type$-1$ (weak) resetting corresponding to continuous PTCs with mean slope $1$ and type$-0$ (strong) resetting corresponding to discontinuous PTCs with mean slope $0$, for a given amplitude $A$. It is evident that the complexity of the phaseless set of the system for each set of parameter values $(\Omega,\eta)$ and the complex accumulation of isochrons in their neighborhood \cite{Guckenheimer, Iso22} affects the shape of the corresponding time crystal. The isochrons and the corresponding phase response curves for a limit cycle of the system (\ref{oisl}) cannot be calculated analytically and their numerical computation is a nontrivial task \cite{Osinga_10}; here, we use an efficient method based on Koopman operator techniques and the computation of Fourier averages along trajectories \cite{Mauroy_12}. In Fig. \ref{tc2}, the time crystal is a quite smooth surface as the phaseless set of Fig. \ref{ps2} consists only of the $1-$dimensional stable manifolds of the saddles and the unstable focus-node. In this case, perturbations along the $x-$axis with amplitude $A\in[0,0.5]$ shift initial conditions on the limit cycle over a $1-$dimensional line segment of the phaseless set. In Fig. \ref{tc1}, the time crystal exhibits abrupt phase transitions due to the intricate shape of the phaseless set consisting of the unstable limit cycle in Fig. \ref{ps1} and the $1-$dimensional stable manifold of the saddle. The role of the phaseless set in the form of the phase response of the system is clearly depicted in the generalized PRCs shown in Fig. \ref{GPRC}.

\section{Synchronization Dynamics and Frequency Comb Shaping}
\subsection{Circle map dynamics}
\begin{figure}  
\begin{center}
    \subfigimg[width=0.45\columnwidth]{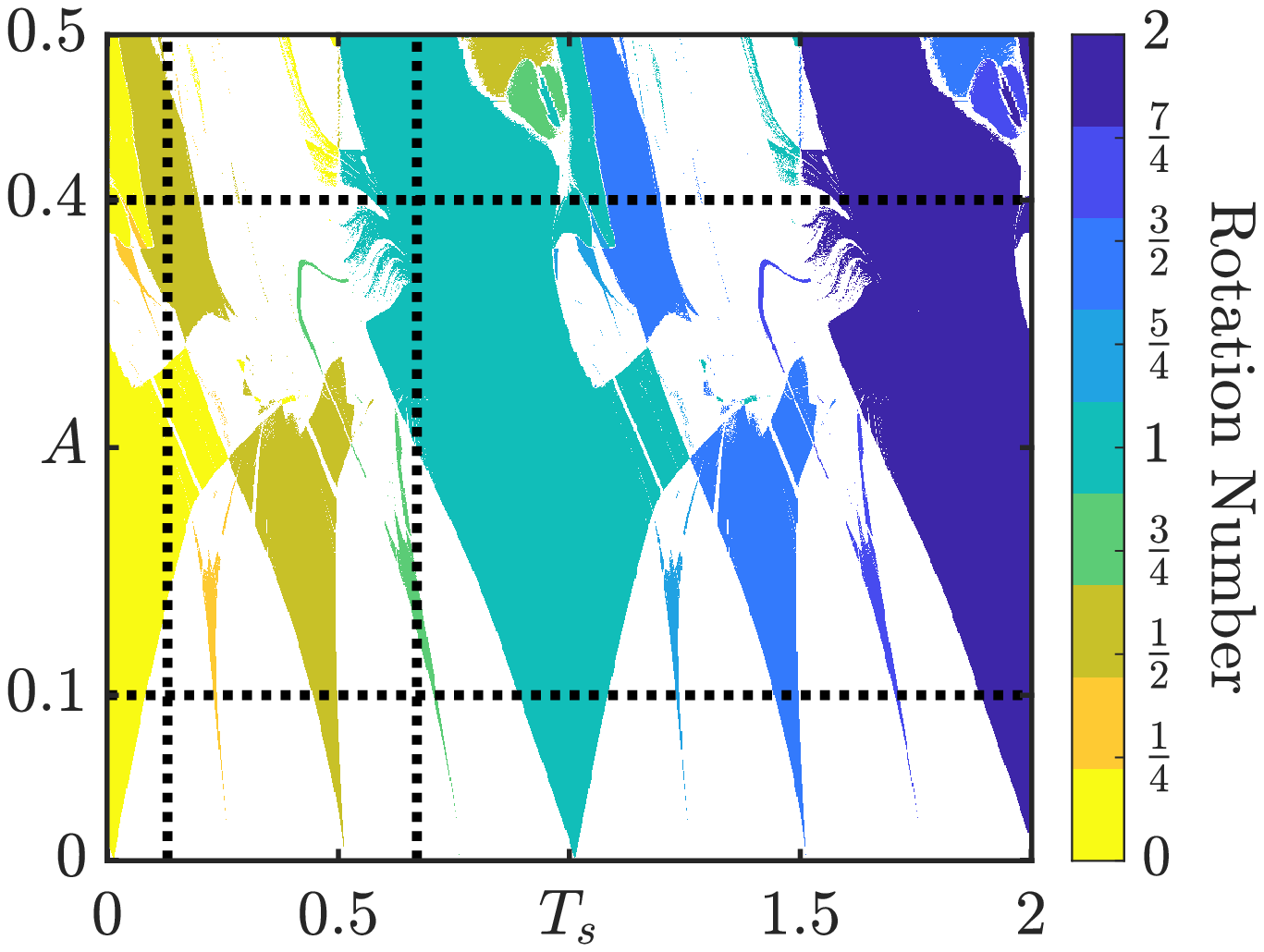}[1\baselineskip]{arn_9}
    \subfigimg[width=0.45\columnwidth]{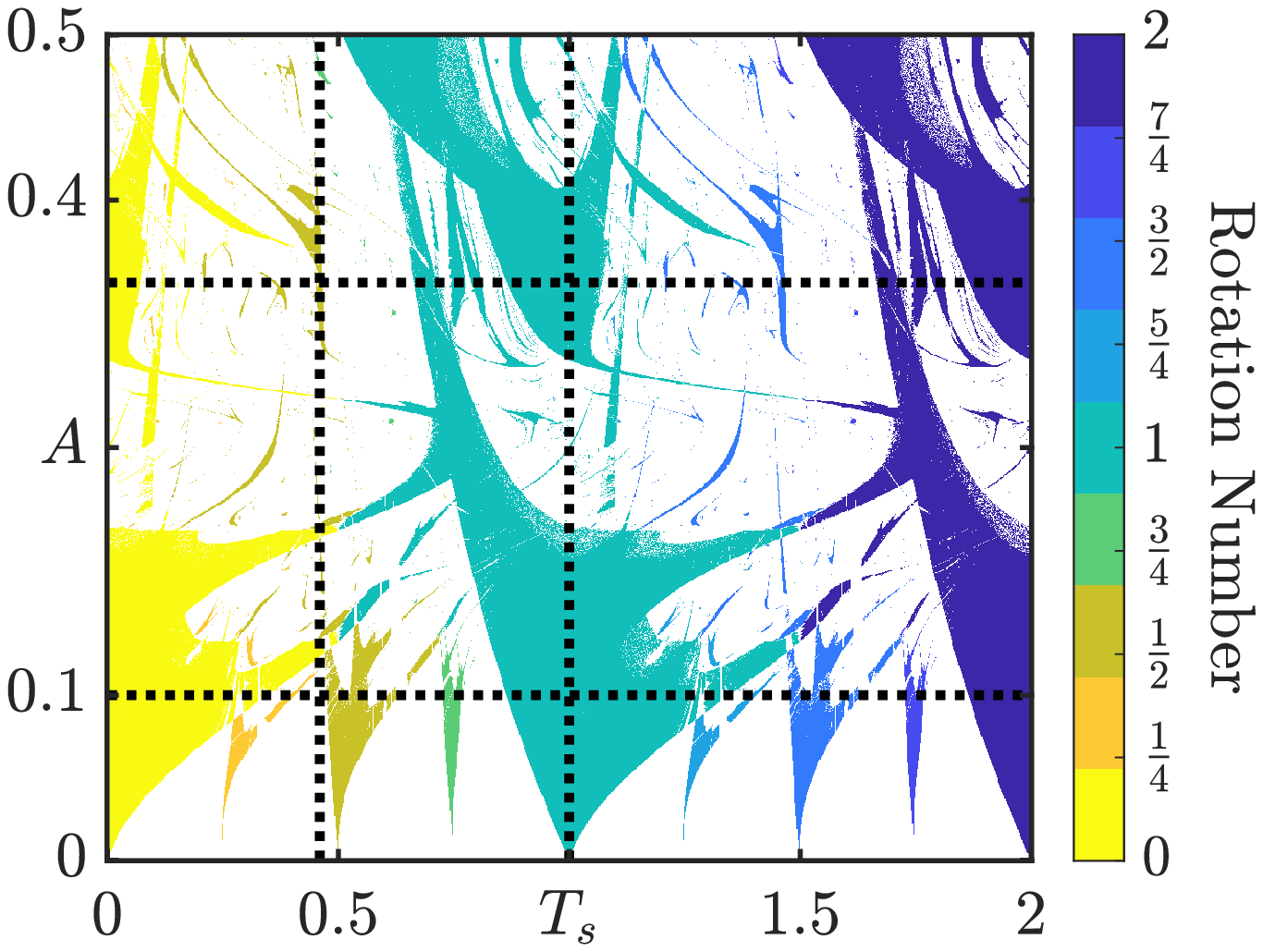}[1\baselineskip]{arn_7}
    \caption{Two-dimensional resonance diagrams for Eq.~(\ref{Poincare}), where (a) corresponds to the time crystal of a limit cycle for $(\Omega, \eta)= (0.06,0.02)$ in Fig. \ref{tc2}, and (b) corresponds to the time crystal of a limit cycle for $(\Omega, \eta)= (0.06,0.0025)$ in Fig. \ref{tc1}. Rotation numbers have been computed as an average of 100 different initial $\theta$-values due to multistability and rounded up to $5$ decimals.} 
    \label{fig:rotation number}
\end{center}
\end{figure}

Phase Response Curves are widely used to describe the behavior of a system operating on a stable limit cycle under periodic forcing \cite{Glass_82, Glass_94, Glass_17, Pikovsky}. Considering a train of periodic pulsatile stimulations of magnitude $A$ delivered with a period $T_s$, that modulates the injected signal of the system (\ref{oisl}),
\begin{equation}
    \eta(t) \rightarrow \eta+A\sum_{n=1}^{m}{\delta{(t-nT_s)}}
    \label{forced system}
\end{equation}
the phase $\theta_{n+1} \in [0, 1)$ at the moment of every stimulus $(n+1)$ is 
\begin{equation}
    \theta_{n+1}=\left[\theta_n+\text{PRC}(\theta_n,A)+T_s\right] \mod 1
    \label{Poincare}
\end{equation} 
where $\theta_n$ is the phase of the system before the stimulus. This equation defines a Poincare mapping of the interval $[0, 1)$ to itself, i.e., a circle map. The essential assumptions in deriving this approximation are: (i) the persistence of the limit cycle's properties under the stimulation, and (ii) the sufficiently large time intervals  between each stimulation or infinite relaxation rate to the limit cycle. Under these assumptions, the dynamics of the forced system (\ref{forced system}) are described by the $1-$dimensional map (\ref{Poincare}). The relaxation interval here is taken equal to $k\cdot T_0$, where $k=100$. The theory of circle maps has immediate applications for the synchronization properties of forced oscillators \cite{Pikovsky}, as a complete qualitative description of the dynamics for any general case of forcing is achievable. Numerical investigation of Eq.~(\ref{Poincare}) leads to the determination of main synchronization regions, where the frequency of the oscillations exactly coincides with that of the external force; that is mathematically expressed as the convergence of the $\{\theta_n\}-$orbit to a fixed point. Higher-order synchronization regions, where the observed frequency is in a rational relation to the external frequency, that is 
\begin{equation}
    \dfrac{\text{observed frequency}}{\text{external  frequency}}=\ \dfrac{p}{q}\ \in\mathbf{Q}
    \label{rotation number}
\end{equation}
can also be identified; here, the $\{\theta_n\}-$orbit converges to a closed orbit. The ratio (\ref{rotation number}) is called \textit{rotation number} $\rho$ and expresses the average increase in $\theta$ per iteration. It depends only on the parameters of the circle map and can be computed for every pair $(\theta_n,A)$ numerically through its definition, that is
\begin{equation}
    \rho=\lim_{n\to\infty}\dfrac{\Theta_n-\Theta_0}{n}
\end{equation}
or by excluding the first (here, $t=500$) transient iterations, i.e.
\begin{equation}
    \rho=\lim_{n\to\infty}\dfrac{\Theta_n-\Theta_t}{n-t}
\end{equation}
where $\Theta$ is the lift of asymptotic phase function $\theta$ to the real axis. In case of multistability it also depends on the choice of the initial conditions. Thus, the rotation number depicted in Fig. \ref{fig:rotation number} corresponds to a mean value of $\rho$ obtained by $100$ different initial $\theta$ conditions and rounded up to $5$ decimals.

\begin{figure}  
\begin{center}
    \subfigimg[width=0.45\columnwidth]{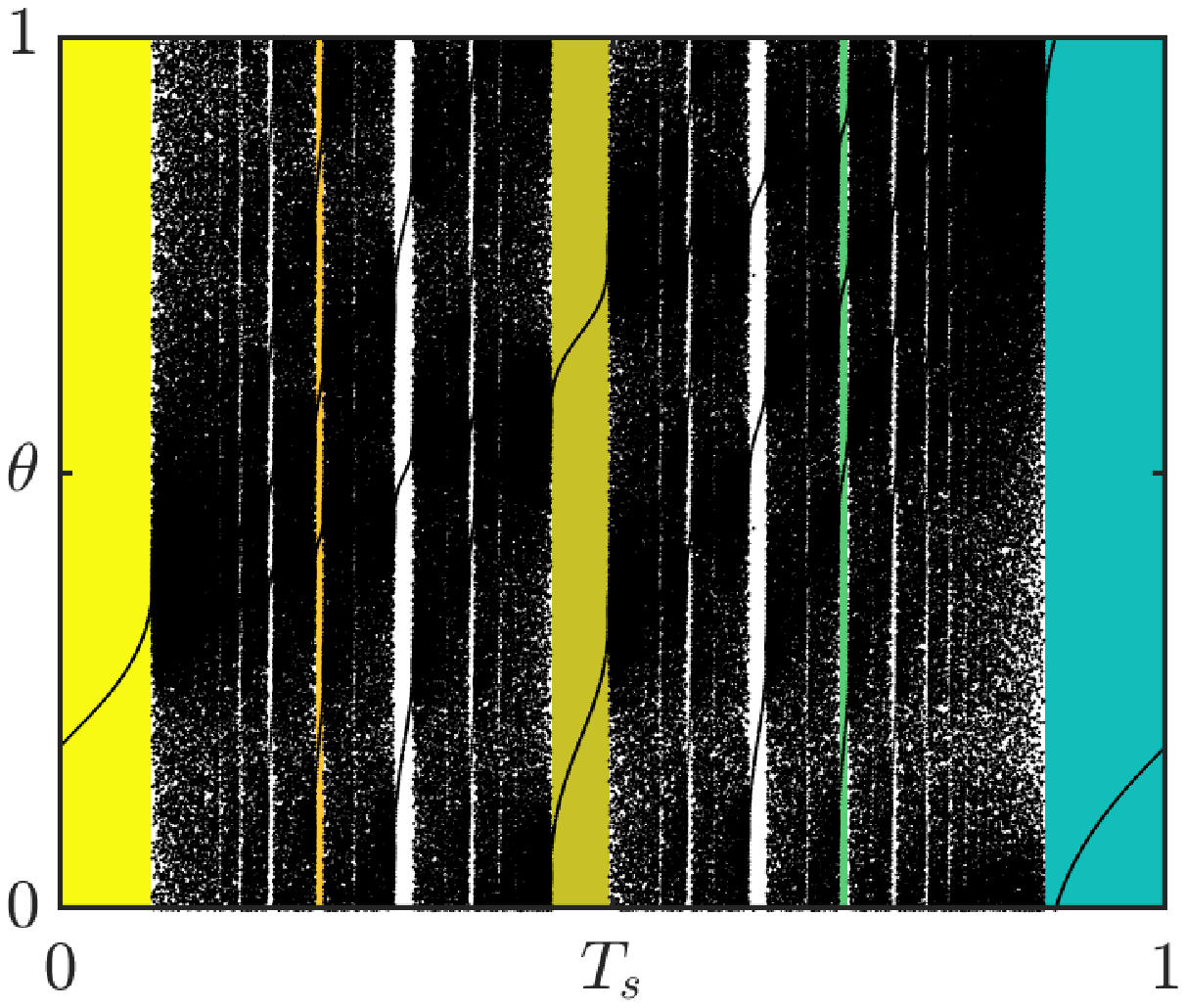}[1\baselineskip]{a=0.1_9}
    \subfigimg[width=0.45\columnwidth]{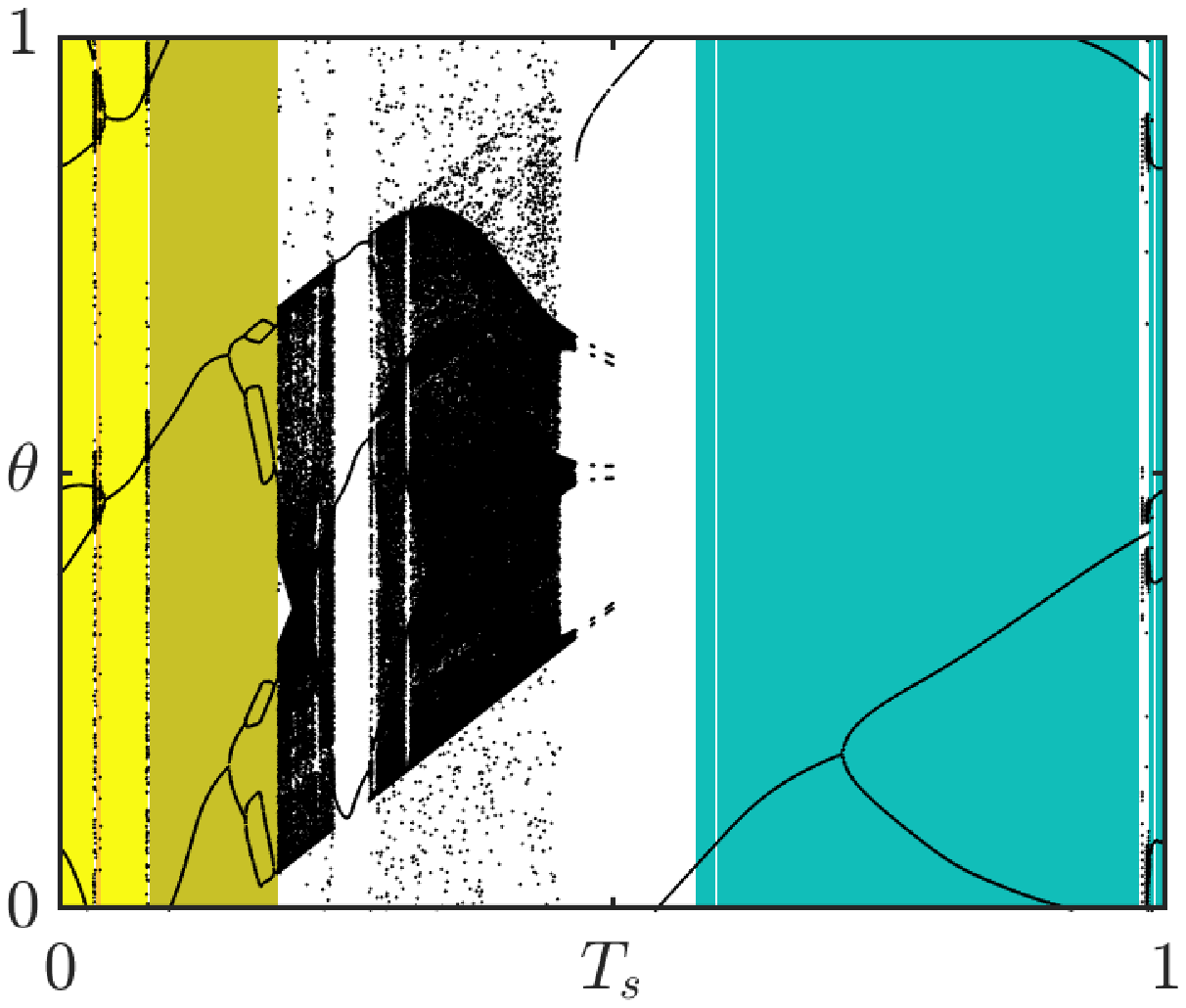}[1\baselineskip]{a=0.4_9}
    \subfigimg[width=0.45\columnwidth]{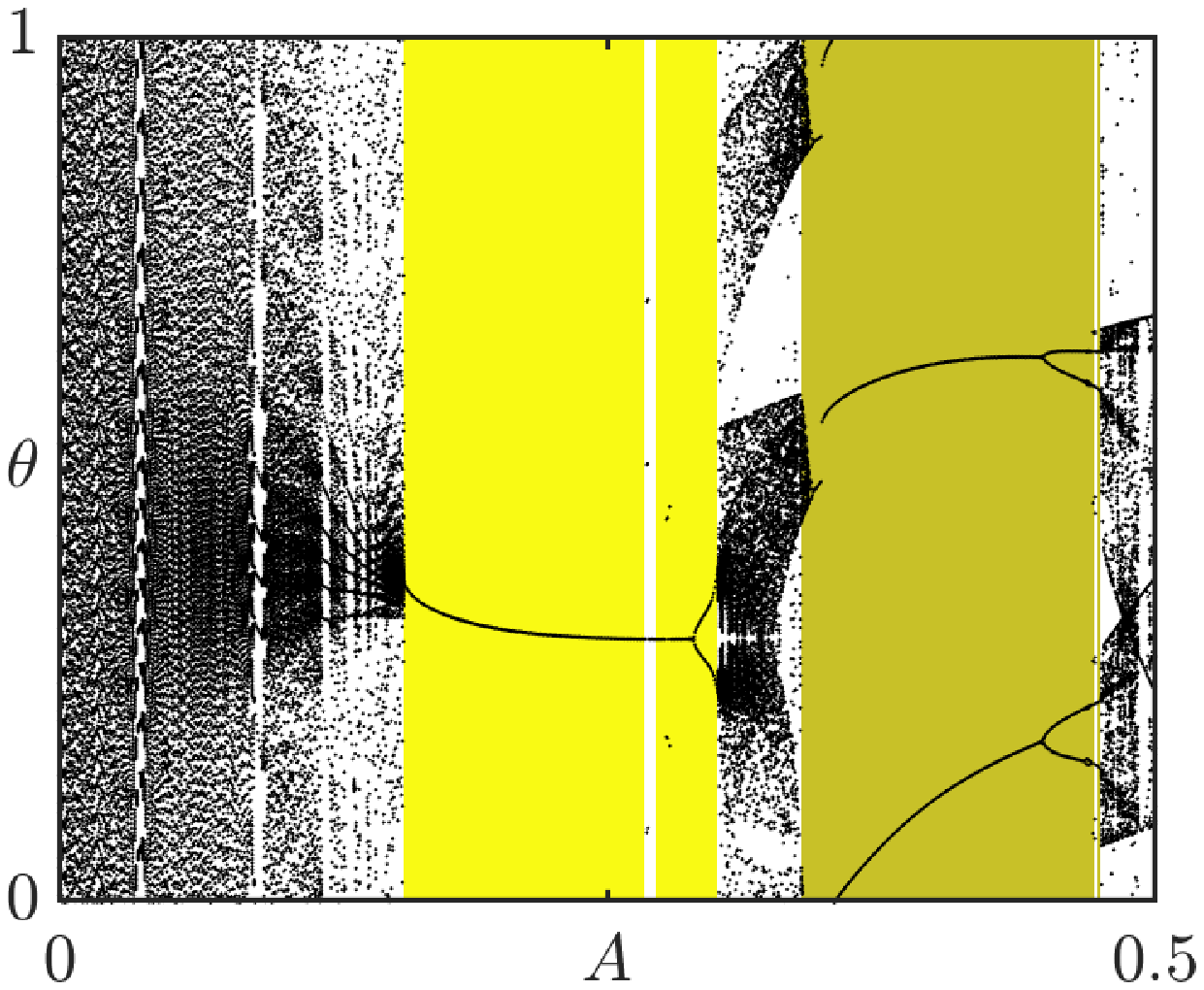}[1\baselineskip]{ts=0.13_9}
    \subfigimg[width=0.45\columnwidth]{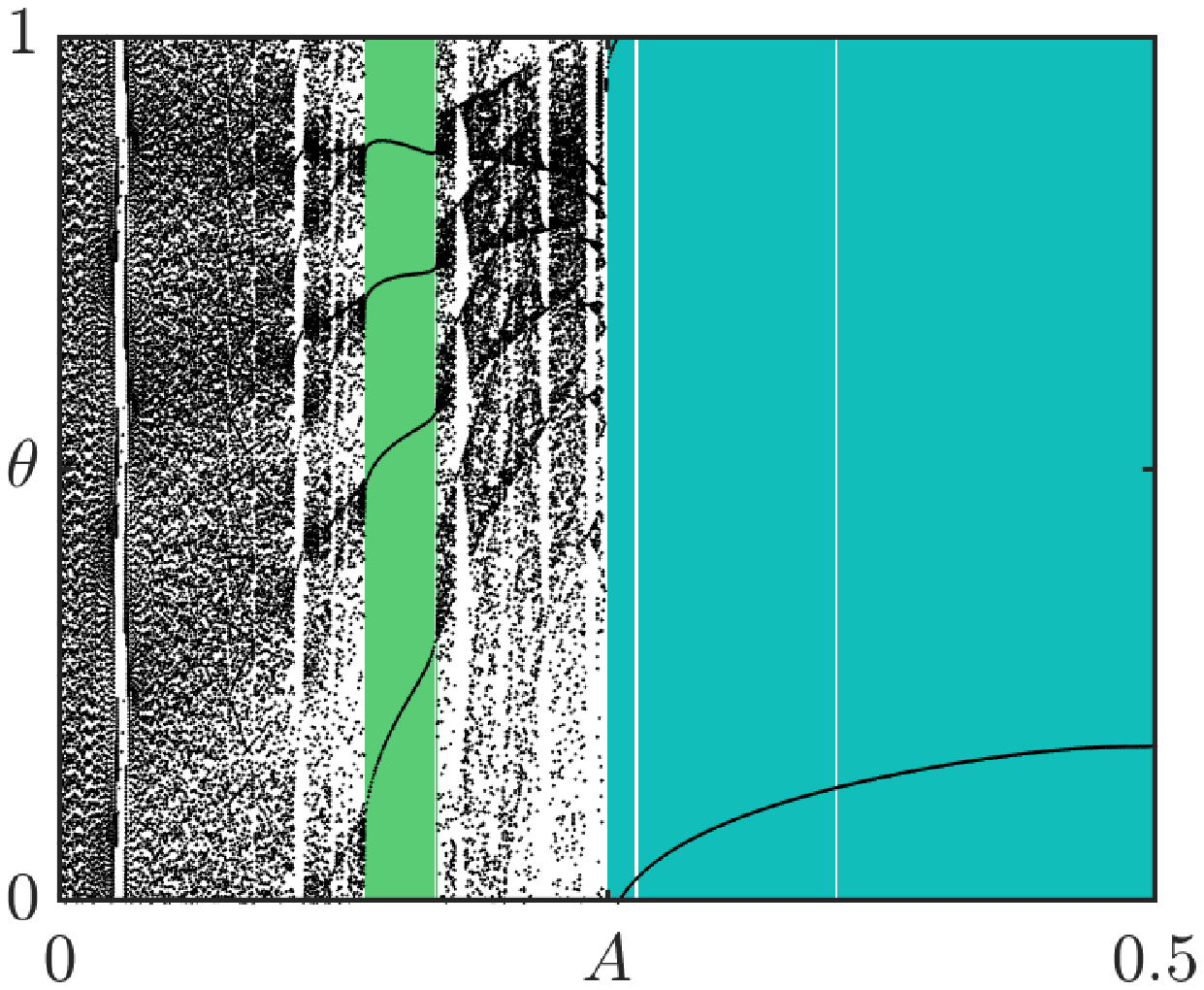}[1\baselineskip]{ts=0.67_9}
    \caption{Bifurcation diagrams for (a) $A=0.1$, (b) $A=0.4$, (c) $T_s=0.13$, and (d) $T_s=0.67$, corresponding to the horizontal and vertical lines in Fig. \ref{arn_9}. Different \{$\theta_n$\} orbits are evolved and their transient parts are excluded. Different colors correspond to various rational rotation numbers according to Fig. \ref{fig:rotation number}.} 
    \label{s2_rn}
\end{center}
\end{figure}
\begin{figure}  
\begin{center}
    \subfigimg[width=0.45\columnwidth]{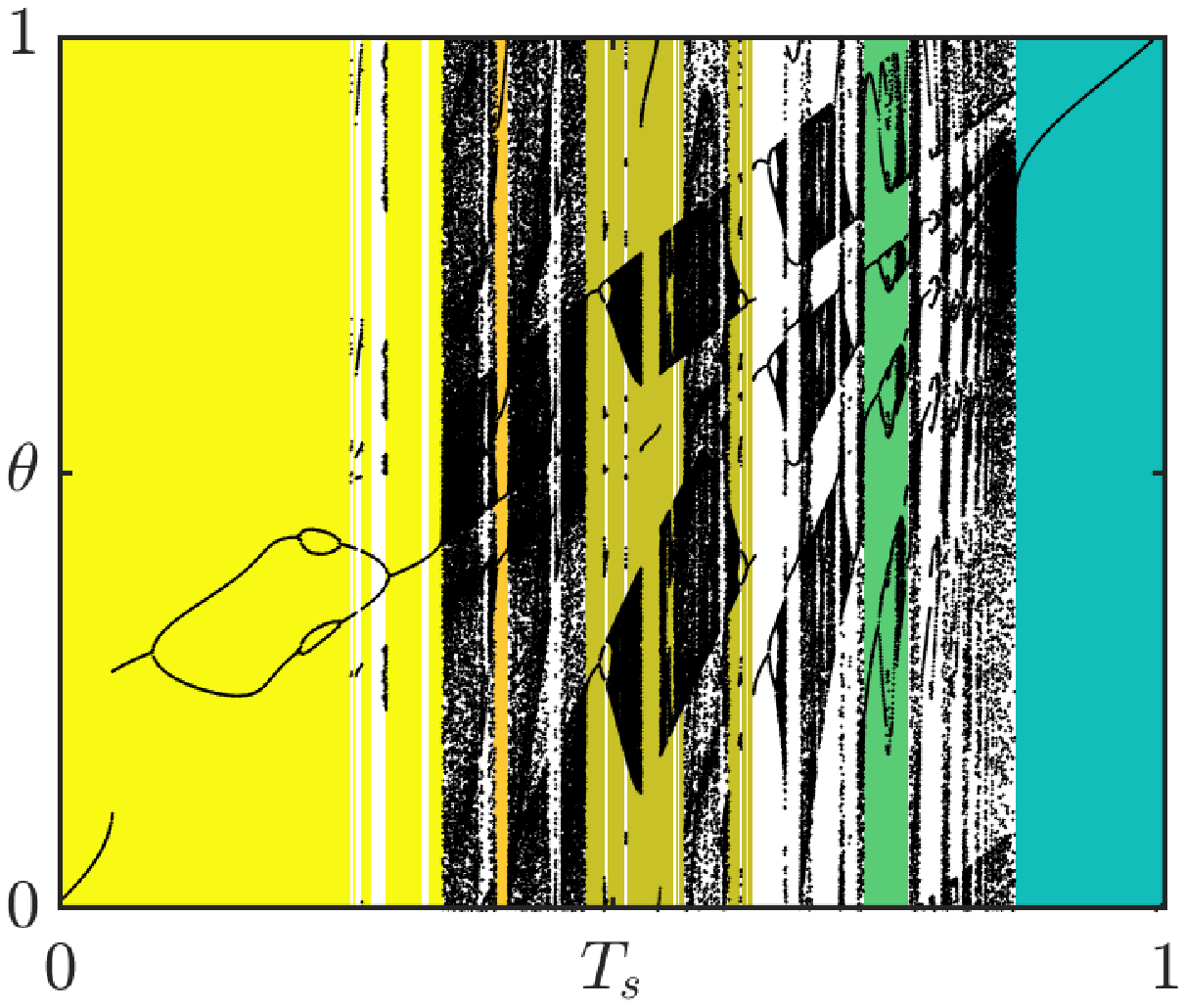}[1\baselineskip]{a=0.1} 
    \subfigimg[width=0.45\columnwidth]{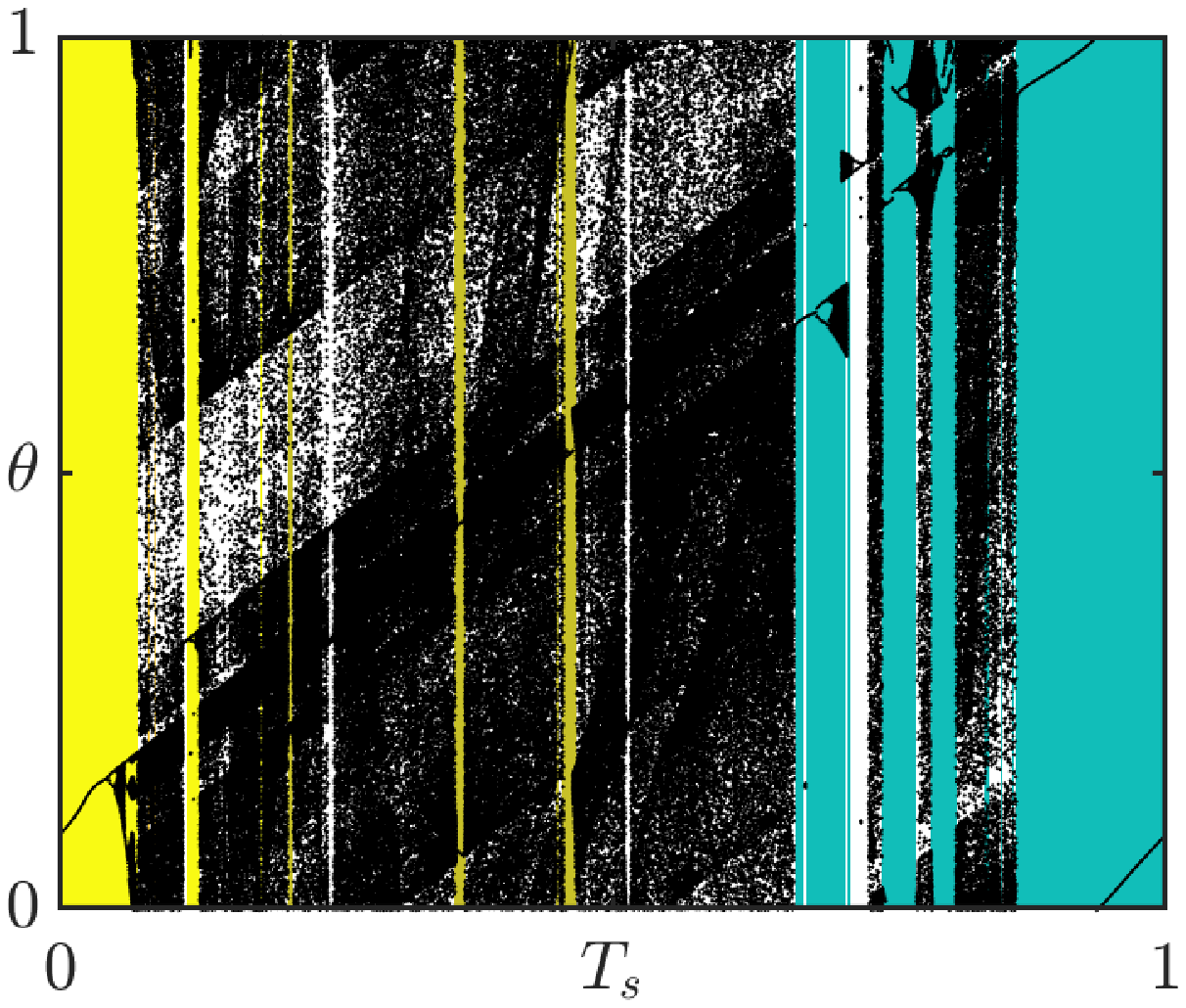}[1\baselineskip]{a=0.4}
    \subfigimg[width=0.45\columnwidth]{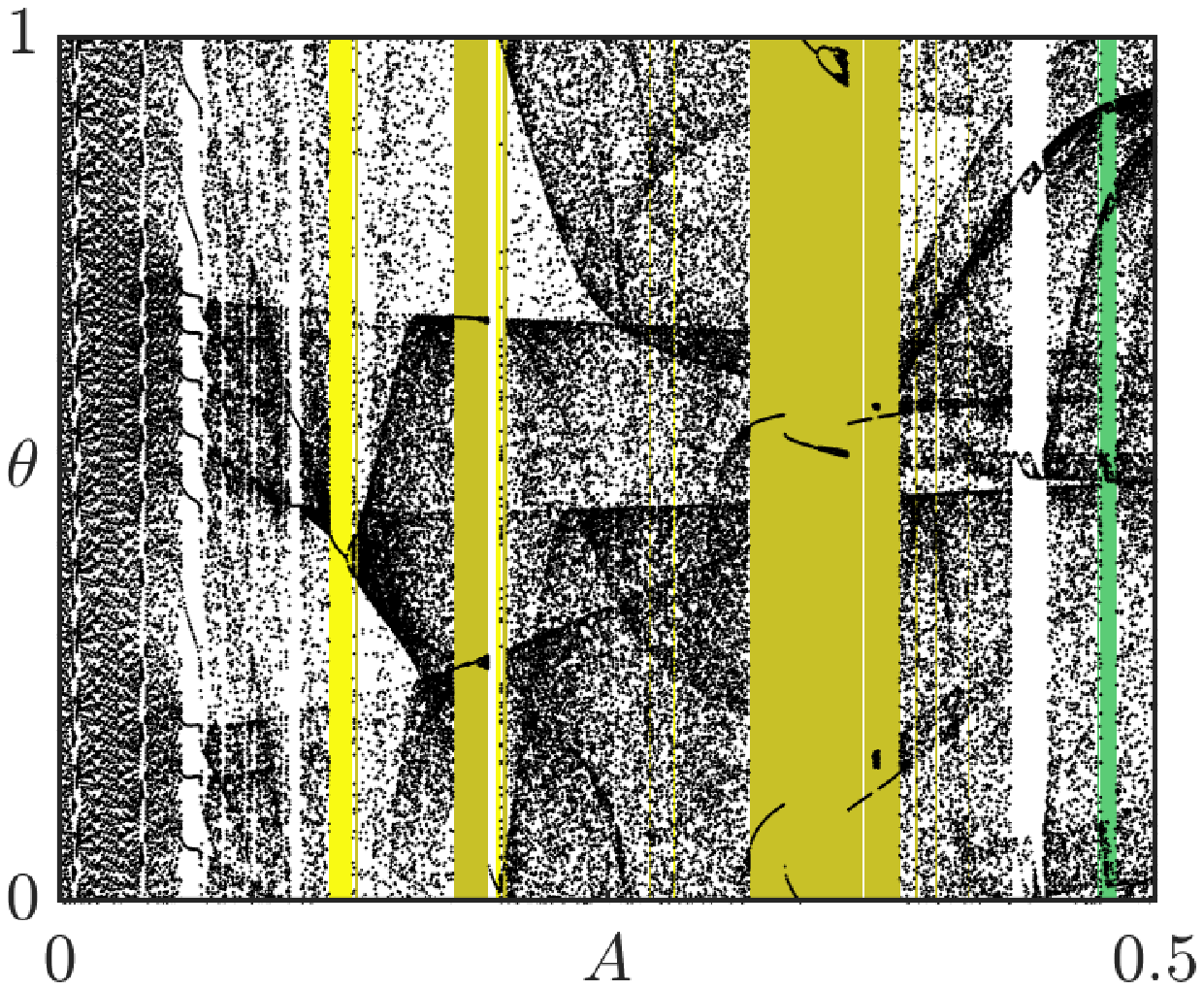}[1\baselineskip]{ts=0.46}
    \subfigimg[width=0.45\columnwidth]{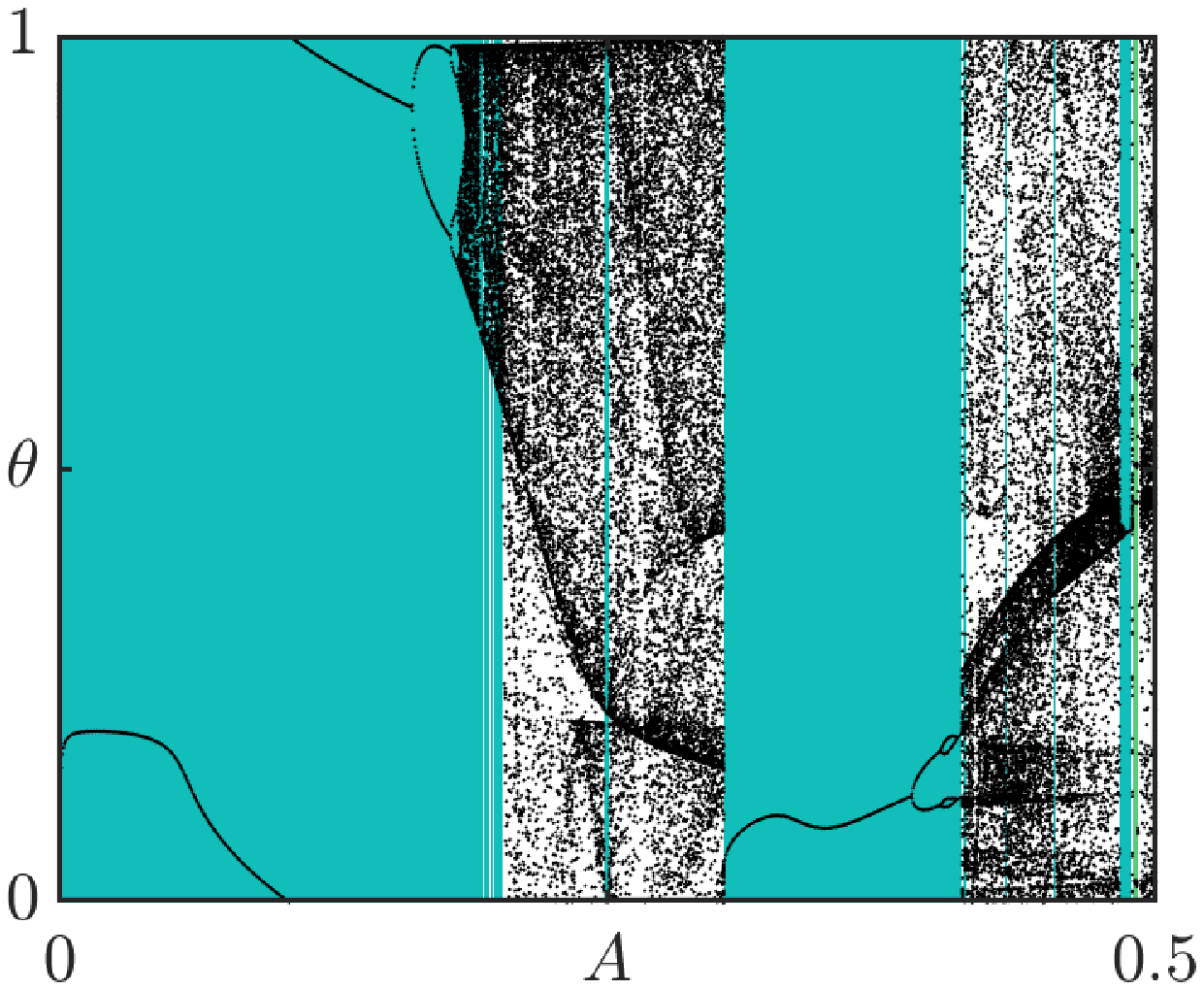}[1\baselineskip]{ts=1}
    \caption{Bifurcation diagrams for (a) $A=0.1$, (b) $A=0.35$, (c) $T_s=0.46$, and (d) $T_s=1$, corresponding to the horizontal and vertical lines in Fig. \ref{arn_7}. Different \{$\theta_n$\} orbits are evolved and their transient parts are excluded. Different colors correspond to various rational rotation numbers according to Fig. \ref{fig:rotation number}. } 
    \label{s1_rn}
\end{center}
\end{figure}

The resulting pattern of locking in circle maps with pulsed forcing is referred to as Arnold-type locking. Arnold provides a detailed understanding of the organization of locking zones in the infinite relaxation limit for low $A-$values. On the plane of the forcing parameters $(T_s,A)$ the synchronization regions that exist are called \textit{Arnold tongues}, and they emanate from rational points on the zero forcing axis. The dynamics within the Arnold tongues and the bifurcations occurring at their boundaries depend strongly on the magnitude of the forcing amplitude with respect to a critical value $A_{cr}$ [$A_{cr}^{(1)}\simeq 0.28725$ and $A_{cr}^{(2)}\simeq 0.13275$ for Fig. \ref{fig:rotation number}(a) and (b)] marking the transition from weak (type$-1$) to strong (type$-0$) resetting. For low to moderate $A$ the dynamics are completely described by the rotation number. For fixed $A$, the boundaries of the Arnold tongues are loci of Saddle-Node bifurcations and outside quasiperiodic motion is observed [Fig. \ref{s2_rn}(a)]. For larger forcing, the dynamics are no longer uniquely determined by the rotation number \cite{Glass_17}.  Different synchronization regions can overlap, thus leading to multistability and Period Doubling bifurcations may take place without any change in the rotation number [Fig. \ref{s2_rn}(b), \ref{s1_rn}(a)] whereas for increasing forcing, Period Doubling route to chaos takes place [Fig. \ref{s1_rn}(b)]. For a fixed forcing period $T_s$ and varying $A$, the bifurcation sequences depend on the way the Arnold tongues are crossed [Figs. \ref{s2_rn}(c-d), \ref{s1_rn}(c-d)].

\subsection{Frequency comb shaping}
\begin{figure*}  
\begin{center}
    \subfigimg[width=0.495\columnwidth]{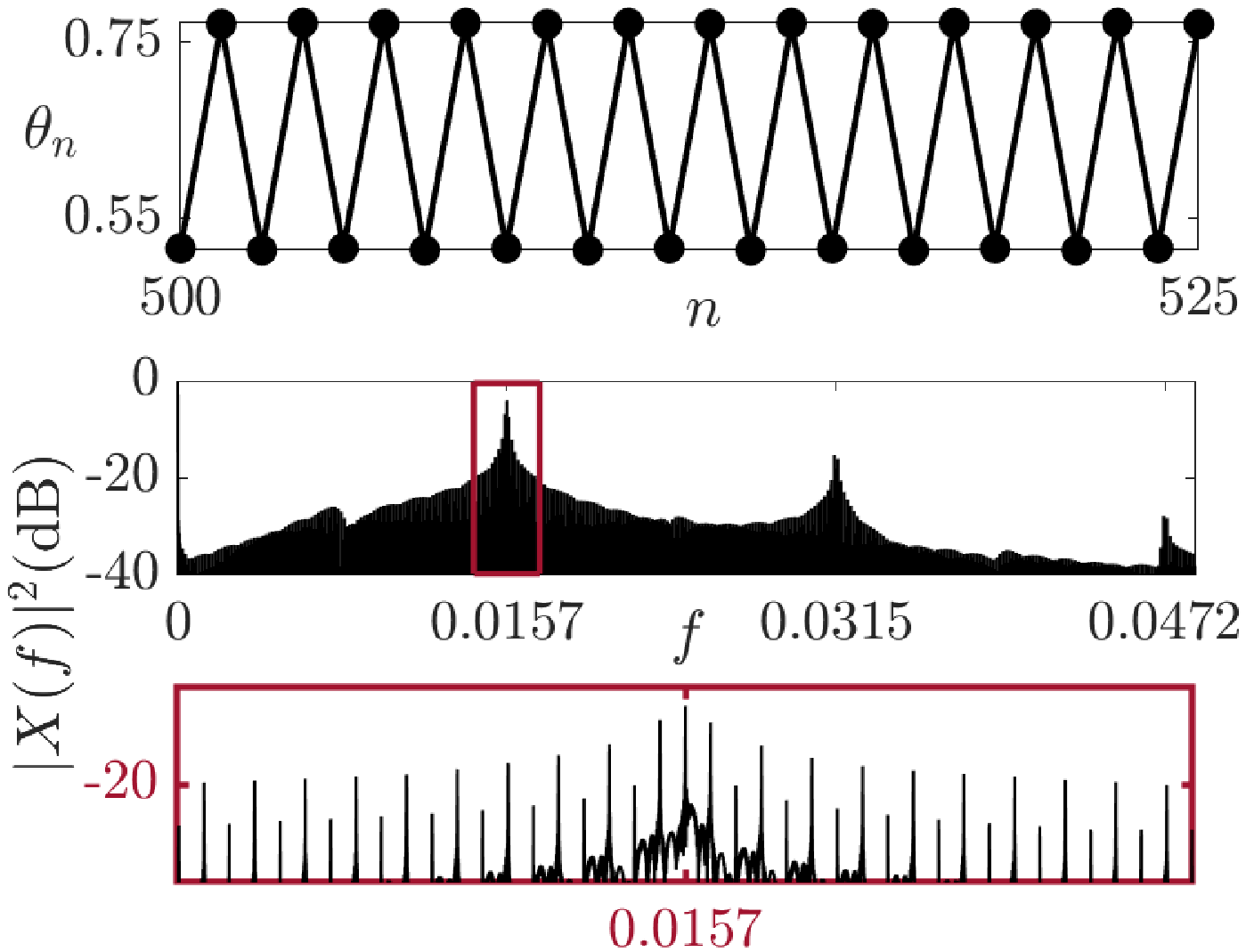}[1\baselineskip]{ex1}
    \subfigimg[width=0.495\columnwidth]{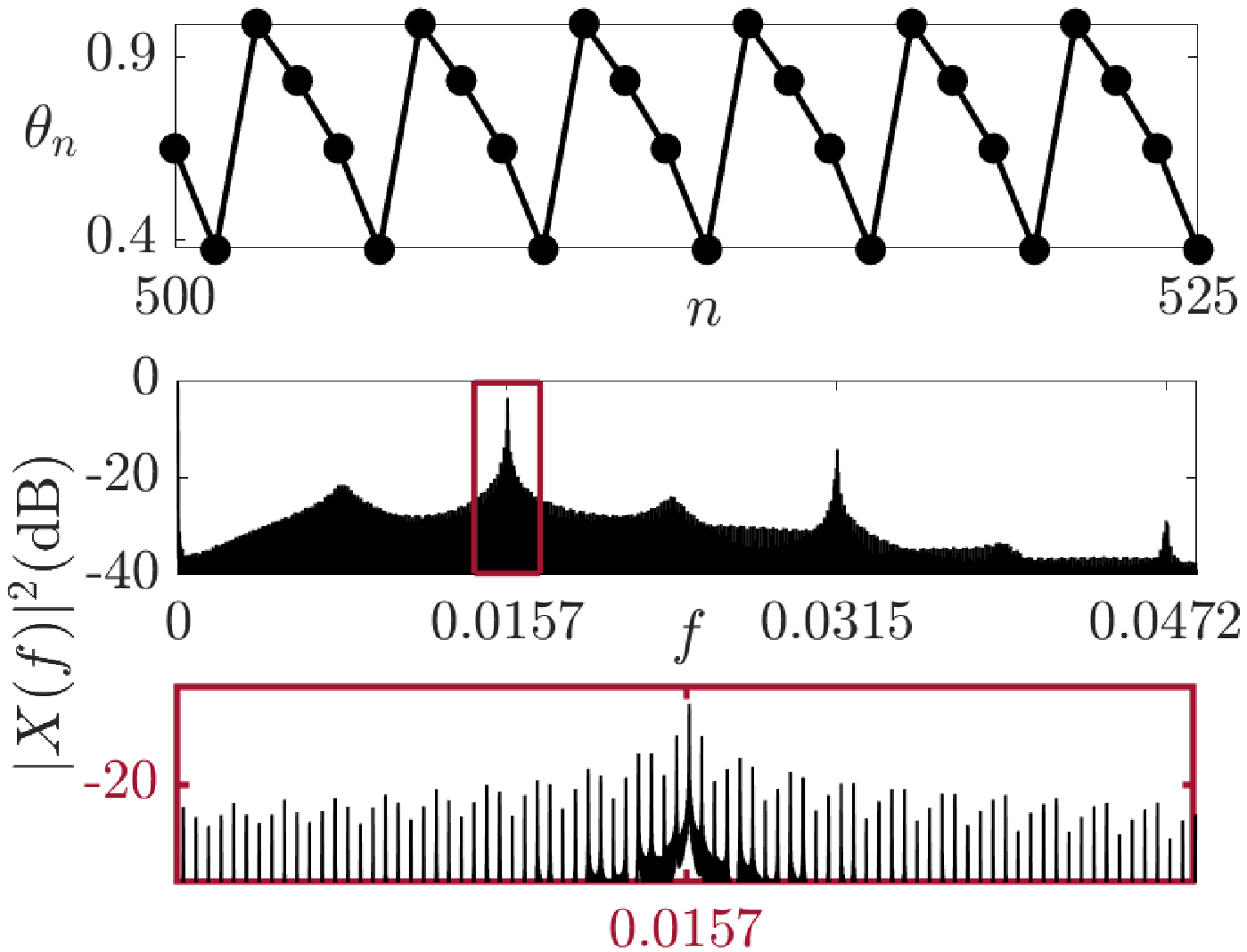}[1\baselineskip]{ex2}
    \caption{Output frequency combs under resonance conditions for a system with $(\Omega, \eta)= (0.06,0.02)$. (top) Circle map orbits: (a) \{$\theta_n$\} orbit with initial condition $\theta_0=0.25$ arising from Eq.~(\ref{Poincare}) for $T_s=0.5$, $A=0.25$. The corresponding rotation number of the phase orbit is $1/4$. (b) \{$\theta_n$\} orbit with initial condition $\theta_0=0.25$ arising from Eq.~(\ref{Poincare}) for $T_s=0.68$, $A=0.13$. The corresponding rotation number of the phase orbit is $3/4$. 
    (middle, bottom) Power spectral density $|X(f)|^2$ calculated from the original system (\ref{oisl}) under the periodic modulation of the injection signal (\ref{forced system}). }
    \label{s2_des}
\end{center}
\end{figure*}

\begin{figure*}  
\begin{center}
    \subfigimg[width=0.495\columnwidth]{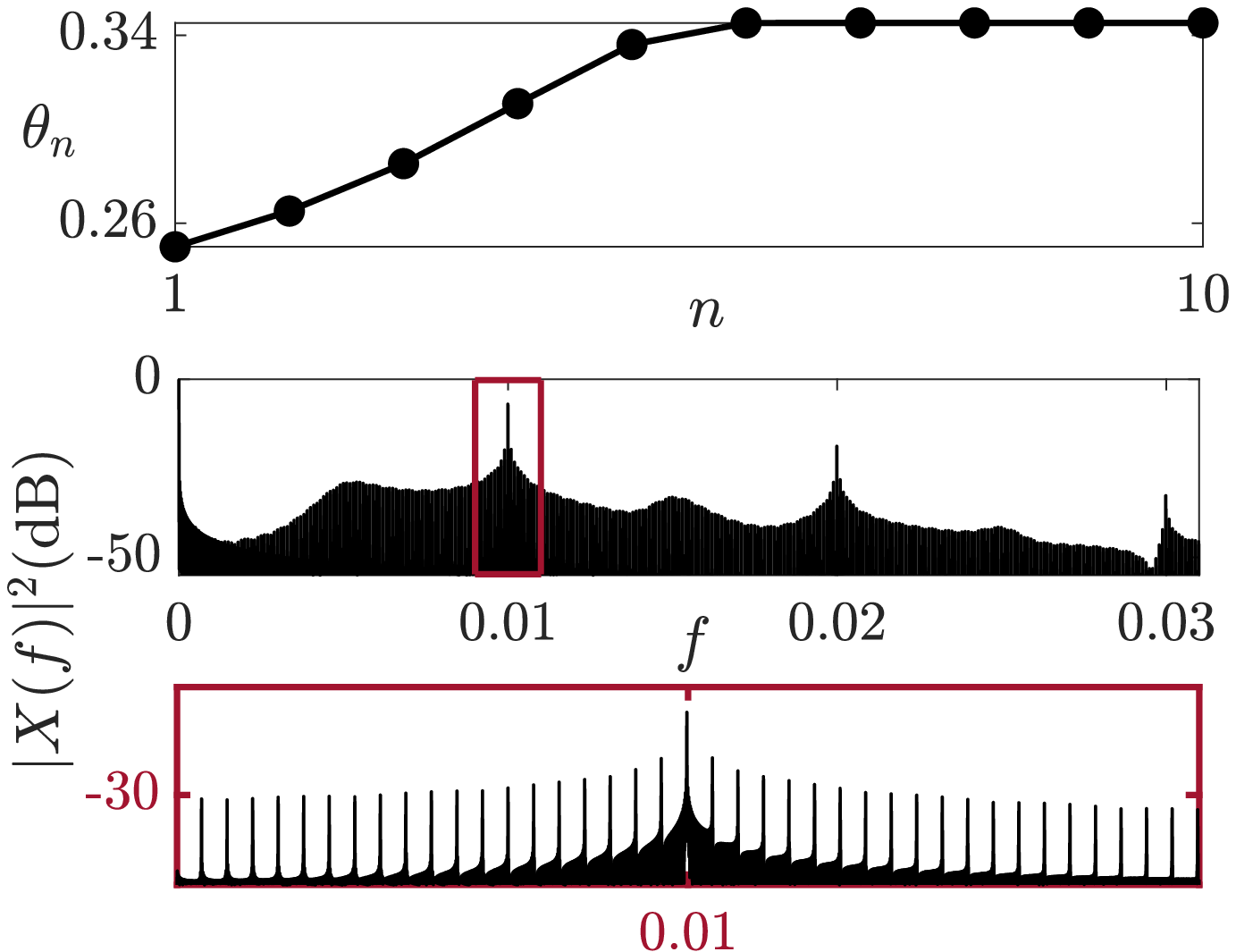}[1\baselineskip]{ex1}
    \subfigimg[width=0.495\columnwidth]{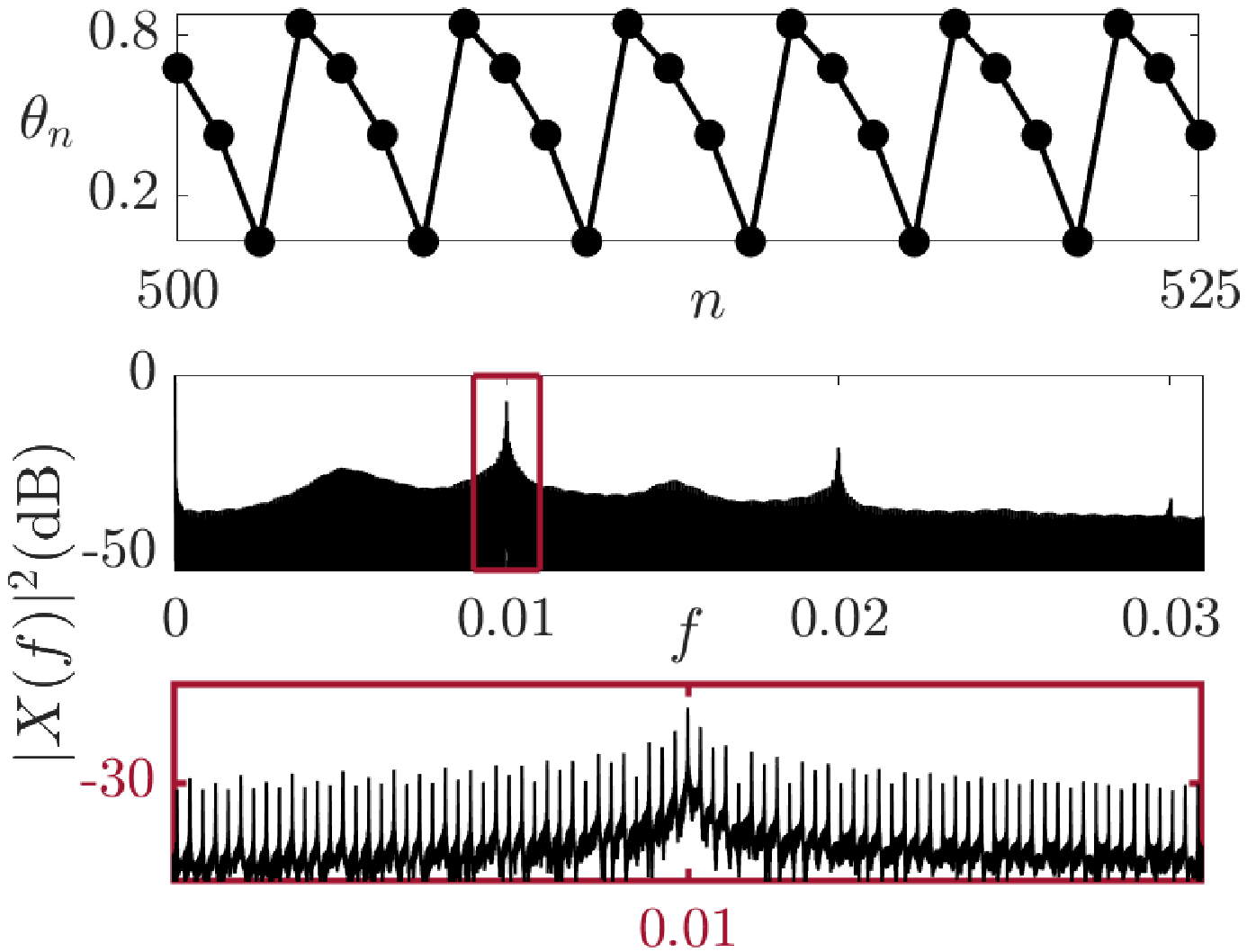}[1\baselineskip]{ex2}
    \caption{Output frequency combs under resonance conditions for a system with $(\Omega, \eta)= (0.06,0.0025)$. (top) Circle map orbits: (a) \{$\theta_n$\} orbit with initial condition $\theta_0=0.25$ arising from Eq.~(\ref{Poincare}) for $T_s=0.05$, $A=0.1$. The corresponding rotation number of the phase orbit is $0$. (b) \{$\theta_n$\} orbit with initial condition $\theta_0=0.25$ arising from Eq.~(\ref{Poincare}) for $T_s=0.75$, $A=0.07$. The corresponding rotation number of the phase orbit is $3/4$. (middle, bottom) Power spectral density $|X(f)|^2$ calculated from the original system (\ref{oisl}) under the periodic modulation of the injection signal (\ref{forced system}). }
    \label{s1_des}
\end{center}
\end{figure*}

The complex dynamics of the circle maps discussed in the previous sections are essentially determined by the Isochrons' structure and the corresponding phase response and Time Crystal of each specific stable limit cycle of the original system (\ref{oisl}) describing the full dynamics of an Optically Injected Laser. This judicious reduction of the three-dimensional original system to a one-dimensional circle map enables the direct connection between the two systems. In Figs. \ref{s2_des} and \ref{s1_des} the periodic solutions corresponding to different rotation numbers of the circle map are shown along with the respective power spectral density of the electric field of the original dynamical system (\ref{oisl}). Conditions for synchronization and periodic solutions of the circle map ensure the periodicity of the output of the original system as shown from the discreteness of the output spectrum. Moreover, the original equidistant, equal-amplitude input frequency comb is drastically transformed. The relative amplitudes of the output spectral lines are modulated according to the nonlinear dynamics of the system. Therefore, selective amplification of the comb lines with the smallest detunings from the limit cycle's spectral content takes place with the major peaks corresponding to harmonics of the limit cycle's frequency. Moreover, the spacing of the output comb spectral lines is determined by the period of the injected frequency comb $T=kT_0+T_s$ as well as by the rational rotation number.

The above-described characteristics suggest that an OISL with modulated injection can perform a processing operation on the input modulated signal. A clear passband filtering effect is exhibited by the resonant character of the synchronization dynamics as shown in Fig. \ref{fig:rotation number}, depending on both the period $T_s$ and the amplitude $A$ of the input signal. In contrast to conventional approaches, the fact that the OISL is a dynamically active device becomes a key issue since it allows for tuning the central passband frequency by varying the frequency of the specific limit cycle within a broad range from 100 MHz to 100 GHz \cite{Herrera2021}. Moreover, its nonlinear character implies an amplitude-dependent selectivity and can drastically transform the periodic input signal, either to another periodic output having a discrete spectrum with controllable spectral line spacing for operating conditions within the passband, or to a chaotic output having a continuous spectrum that can be useful for chaos-based applications \cite{Sciamanna2015}.

\section{Summary and Conclusions}
The concept of a Time Crystal has been introduced for the study of tunable photonic oscillators, consisting of Optically Injected Semiconductor Lasers, driven by frequency combs. The time crystals have been shown capable of containing all the information for the phase response of the underlying stable limit cycle of the system and enabling the judicious dynamical reduction of the system under a periodic pulsatile input to a one-dimensional circle map. The form of the phase response curve directly dictates the synchronization properties of the system such as the frequency range for stable locking. The circle map has been studied with respect to its complex synchronization dynamics in terms of resonance diagrams, showing Arnold tongues, and bifurcation diagrams, exhibiting periodic and chaotic dynamics. The dynamics of the circle map has been shown accurate in describing the dynamics of the original system and providing conditions for the existence of output frequency combs. Moreover, a remarkable tunability of the system's output, depending on the forcing amplitude and frequency and dictated by the specific characteristics of each time crystal, has been presented. These features suggest that the presented theoretical investigations based on the concepts of Isochrons, Phase Response Curves and Time Crystals can have potential for practical applications related to photonic signal-processing units.

\newpage
\clearpage
\nocite{*}
\bibliography{aipsamp}

\newpage

\clearpage

\clearpage

\clearpage

\clearpage

\end{document}